\documentclass[dvips,11pt]{article}

\usepackage[latin1]{inputenc}  

\usepackage{mathrsfs}
\usepackage[pdftex]{graphicx}
 \usepackage{graphicx}

\def\p{\partial}

\def\th{{\vartheta}}

\def\no{\noindent}
\def\sm{{\smallskip}}
\def\me{{\medskip}}
\def\bx{{\mathbf x}}
\def\bv{{\mathbf v}}

\def\bq{{\mathbf q}}
\def\f{{\mathbf f}}

\def\pb{\, .}
\def\pa{\, \cdotp}
\def\vb{\, ,}
\def\pa{\cdotp}
\def\va{\raise 2pt\hbox{,}}

\def\bull{{\vrule height.9ex width.8ex depth.1ex}}

\def\p{\partial}
\def\ve{{\varepsilon}}

\def\a{{\alpha}}
\def\b{{\beta}}
\def\si{{\sigma}}

\def\cA{{\mathcal{A}}}
\def\cB{{\mathcal{B}}}
\def\cC{{\mathcal{C}}}

\def\cJ{{\mathcal{J}}}

\def\bfR{\mbox{\boldmath$R$}}

\def\p{\partial}

\def\no{\noindent}
\def\sm{{\smallskip}}

\def\bx{{\mathbf x}}
\def\bv{{\mathbf v}}

\def\f{{\mathbf f}}

\def\pb{\, .}
\def\pa{\, \cdotp}
\def\vb{\, ,}
\def\pa{\cdotp}
\def\va{\raise 2pt\hbox{,}}

\def\bq{{\mathbf q}}
\def\f{{\mathbf f}}

\def\th{{\theta}}

\def\RR{{\mathbf R}}

\def\p{\partial}
\def\ve{{\varepsilon}}

\def\a{{\alpha}}

\def\cA{{\cal A}}
\def\cB{{\cal B}}
\def\cC{{\cal C}}

\def\cJ{{\cal J}}

\def\bfR{\mbox{\boldmath$R$}}

\def\p{\partial}

\def\no{\noindent}
\def\sm{{\smallskip}}

\def\bx{{\mathbf x}}
\def\bv{{\mathbf v}}

\def\f{{\mathbf f}}

\def\pb{\, .}
\def\pa{\, \cdotp}
\def\vb{\, ,}
\def\pa{\cdotp}
\def\va{\raise 2pt\hbox{,}}

\def\bq{{\mathbf q}}
\def\f{{\mathbf f}}

\def\th{{\theta}}

\def\RR{{\mathbf R}}

\def\p{\partial}
\def\ve{{\varepsilon}}

\def\a{{\alpha}}

\def\cA{{\cal A}}
\def\cB{{\cal B}}
\def\cC{{\cal C}}

\def\cJ{{\cal J}}

\newtheorem{thm}{Theorem}[section]

\newtheorem{lemma}[thm]{Lemma}

\newtheorem{theorem}[thm]{Theorem}

\newcommand{\be}{\begin{equation}}
\newcommand{\ee}{\end{equation}}

\begin{document}

{\bf \title { From the Micro-scale to Collective Crowd Dynamics.}}

\author{Nicola  Bellomo\thanks{Department of Mathematics, Politecnico Torino
Corso Duca degli Abruzzi 24, 10129 Torino, Italy,
 ({\tt nicola.bellomo@polito.it}). Partially supported by MIUR, Italian Minister for University and Research}
\and Abdelghani Bellouquid\thanks{University Cadi Ayyad, Ecole Nationale des Sciences Appliqu\'ees, Safi, Morocco,({\tt bellouq2002@yahoo.fr}). Supported by Hassan II Academy of Sciences and Technology (Morocco), Project ``M\'ethodes math\'ematiques et outils de mod\'elisation
et simulation pour le cancer''} \and Damian Knopoff\thanks{Facultad de Matem\'atica, Astronom\'ia y F\'isica,
 University of Cordoba, CIEM-CONICET C\'ordoba, Argentina,({\tt damianknopoff@gmail.com}).}}

\date{}
\maketitle

\begin{abstract}

 This paper deals with the kinetic theory modeling of crowd dynamics with the aim of showing how the dynamics at the micro-scale is transferred to the dynamics of collective behaviors. The derivation of a new model is followed by a qualitative analysis of the initial value problem. Existence of solutions is proved for arbitrary large times, while  simulations are developed  by computational schemes based on splitting methods, where the transport equations  treated by finite difference methods for hyperbolic equations. Some preliminary reasonings toward the modeling of panic conditions  are proposed.
\end{abstract}


\maketitle

\section{Introduction}\label{INTRO}

The modeling of pedestrian crowd dynamics can be developed at different representation scales, namely micro- and macro-scales with the intermediate approach of the kinetic theory, where the dependent variable is a probability distribution over the micro-state of pedestrians. The interested reader is addressed to the review paper \cite{[BPT12]}, where the existing literature is reviewed and critically analyzed focusing at different modeling scales.  Namely,  at the micro-scale by ordinary differential equations, see among others \cite{[HEL91],[HJA07],[HM09]};  at the macro-scale obtained by the classical approach by conservation equations, among others \cite{[CC08],[MRS10]}; or by  stochastic models related to evolving probability measure \cite{[PT11]}.  The interplay between individual-based and macroscopic models is studied in \cite{[CPT11]}. Additional bibliography and conceptual links with the modeling of vehicular traffic are proposed in \cite{[BD11]}. A detailed analysis of empirical data is delivered by various papers \cite{[BW06],[RWTH11],[SS11]} among others.

This present paper specifically refers to \cite{[BB11]}, where the main hint, further stressed in  \cite{[BPT12]}, is that the modeling approach should retain, as far as it is possible, the complexity features of crowds to be viewed as a living, hence complex, system. This paper presents an introduction to modeling where pedestrians move with only velocity modulus and change their velocity directions both due to interactions with other pedestrians and to the search of their specific target, for instance the exit zone. This present paper  shows how the dynamics at the micro-scale of pedestrians  is transferred to the dynamics of collective behaviors. Modeling and simulations include interactions between pedestrian that move toward different targets, while their dynamics can be induced, by signals from the outer environment to an exit direction. A qualitative analysis of the initial value problem, and some sample simulations are presented.

The approach developed in this paper is based on the methods of the kinetic theory of active particles,  which  shows the ability to retain various complexity features. This approach has been applied to the modeling of several living systems as documented, among others, in the modeling of vehicular traffic \cite{[BB11B],[BDF12]}, swarms \cite{[BS12]}, social competition \cite{[ABE08]}, and migration phenomena \cite{[DK12]}.

The contents are proposed through four additional sections. In details,  Section 2 presents the mathematical model obtained as a development of that given in \cite{[BB11]} to include the dynamics on the velocity modulus and interactions of pedestrians that move toward different directions \cite{[ADM11]}. Section 3 introduces the problem of existence of solutions to the initial value problem in unbounded domains and announces the existence theorems for arbitrary large times. The technical proofs are developed in Sections 4 and 5. Simulations obtained by application of the splitting method, are presented in Section 6 with a detailed computational analysis of the dynamics of the moving boundary of the domain occupied by the crowd.  Finally, the last section first critically analyzes the use of the kinetic theory approach and subsequently looks ahead to research perspectives focusing on the modeling panic conditions.


\vskip.5truecm

\section{A Mathematical Model}

\begin{figure}[h!]
\bigskip
\begin{center}
  \includegraphics[width=0.55\textwidth]{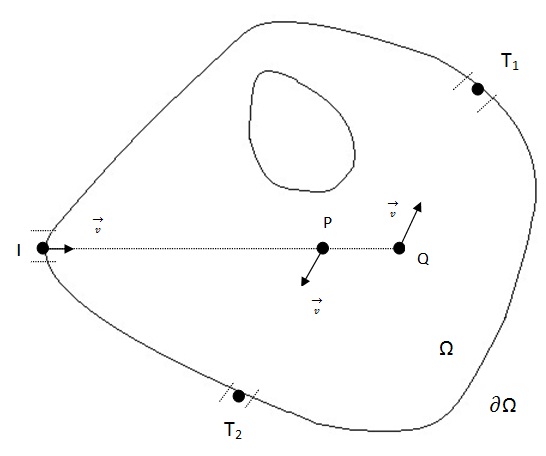}
 \end{center}
  \caption{ --  Geometry of the domain $\Omega$: Pedestrians enter from $I$ and  can move towards two different targets: pedestrians at $P$ and at $Q$ move to  $T_2$ and $T_1$, respectively. }\label{fig:geometry_domain}
\end{figure}

Let us consider a large system of pedestrians moving in a bounded domain $\Omega \subseteq \bfR^2$, which may include, as shown in Figure \ref{fig:geometry_domain}, internal obstacles.
The boundary of $\Omega$ is denoted by $\p \Omega$. The modeling approach also refers to unbounded domains in two space dimensions. Pedestrians can be subdivided into two different groups, moving toward two different targets, denoted by $T_1$ and $T_2$, respectively.

\begin{figure}[h!]
\bigskip
\begin{center}
\begin{tabular}{cc}
\includegraphics[width=0.15\textwidth]{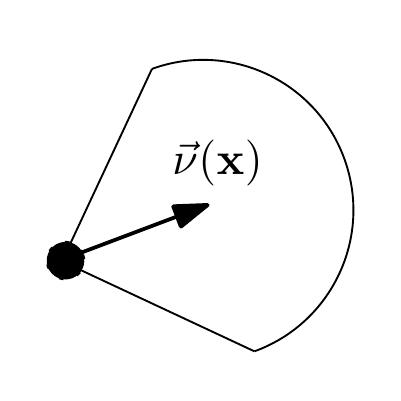} &
\hspace*{-0.4cm}\includegraphics[width=0.55\textwidth]{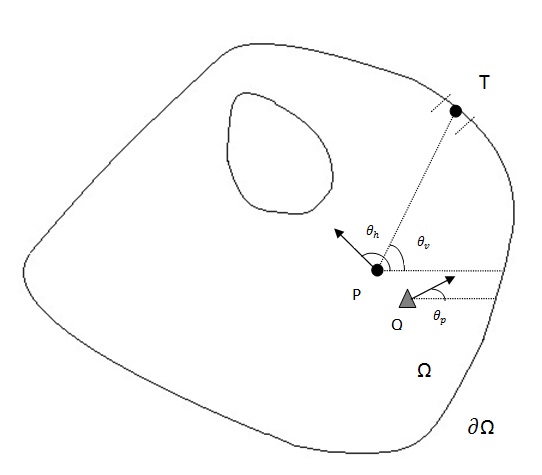}
\\
(a) & (b)
\end{tabular}
\end{center}
\caption{(a) Interaction domain for an individual in the position $\bx$. (b) Interaction between a candidate particle in the position $P$ moving in the direction $\theta_h$ and a field particle in $Q$ moving in the direction $\theta_p$; $\theta_\nu$ is the angle from $P$ to the target $T$.}
\label{fig: visibility_interactions}
\end{figure}

A preliminary step of the derivation of the model, following \cite{[BB11]}, consists in the assessment of the representation of the system according to the
methods of the generalized kinetic theory. This means selecting the microscopic state of the individual state of the pedestrians and of the probability distributions over such state. Two distributions are considered, corresponding to the aforesaid groups of pedestrians pursuing different objectives, which are the dependent variables of the equation modeling the overall dynamics of the system.

Bearing all above in mind, the microscopic state can be defined by the variables: position  $\bx = \{x, y\} \in \Omega \subseteq \bfR^2$, and speed $\bv \in D_v \subseteq \bfR^2$. Dimensionless coordinates are used by  dividing space coordinates by the largest dimension $\ell$ of $\Omega$, while the velocity modulus $v$ is referred to the largest admissible velocity $v_\ell$ that can be reached by a speedy pedestrian in optimal conditions of the ambient where the crowd moves.

Polar coordinates with discrete values are used for the velocity variable $\bv = \{v, \th\}$, where $v$ is the velocity modulus and $\th$ identifies the velocity directions with respect to the $x$-axis:
$$I_\th = \{\th_1 = 0\vb \ldots \vb \th_i \vb \ldots \vb \th_n =  \frac{n}{n-1} 2\pi\},$$

$$\quad I_v = \{v_1 = 0\vb \ldots \vb v_j \vb \ldots \vb v_{m}  = 1\}.$$


Out of this discrete velocity representation some specific cases can be considered. For instance the case of two different groups, labeled with the superscript $\sigma = 1, 2$, which move towards two different targets. Moreover, if the activity variable is assumed to be homogenously distributed for all pedestrians, the corresponding  representation is   as follows:
\begin{equation}\label{representation}
f^\si (t, \bx, \bv, u) = \sum_{i=1}^n \sum_{j=1}^m f_{ij}^\si (t, \bx) \,  \delta (\th - \th_i) \otimes \delta (v - v_j) \otimes \delta (u - u_0)\vb
\end{equation}
where $f_{ij}^\si(t,\bx) = f^\si(t,\bx,\th_i, v_j, u_0)$ corresponding, for each group $\si = 1,2$, to the $ij$-particle, namely to the pedestrian moving in the direction $\th_i$ with velocity $v_j$.

Macroscopic quantities are obtained by weighted sums. In particular, the number density and flow of pedestrians are, respectively,  given by:
\begin{equation}
\rho (t, \bx)  =  \sum_{\si=1}^2 \rho^\si (t, \bx) = \sum_{\si=1}^2 \sum_{i=1}^n \sum_{j=1}^m f_{ij}^\si (t, \bx)  \vb
\end{equation}
and
\begin{equation}
 \bq(t, \bx) =  \sum_{\si=1}^2  \bq^\si (t, \bx)  =  \sum_{\si=1}^2\sum_{i=1}^{n} \sum_{j=1}^m  \left(v_j\, \cos \th_i\,, v_j \, \sin \th_i \,\right) f_{ij}^\si (t, \bx).
\end{equation}

\sm \no {\bf Remark 2.1.}
The use of discrete space  microscopic states  was introduced in \cite{[DT07]}  and \cite{[CDF07]} to model the granular behavior of vehicular traffic, namely it is observed in \cite{[DA95]}  that the number of vehicles is not large enough to justify the assumption of
a continuous distribution function. The same reasonings are valid also in the case of pedestrians in a crowd.

\sm \no {\bf Remark 2.2.}
The more general case, where the activity variable is heterogeneously distributed and is modified by interactions is critically analyzed in the last section focusing on the modeling of panic conditions.
\sm

The derivation of models needs to be related to a mathematical structure suitable to retain, as far as possible, the complexity  of crowds dynamics. Accordingly, the following qualitative features of the dynamics are taken into account:

\sm \no i)  Pedestrians have a  visibility zone $\Lambda = \Lambda(\bx)$, see Figure \ref{fig: visibility_interactions}(a), which does not coincide with the whole domain $\Omega$ due to the limited visibility angle of each individual. The specific geometry of $\Omega$ can reduce the visibility zone for pedestrians near walls.

\sm \no ii) Interactions, at each  at time $t$, involve: the {\bf test} $(ij)$-particle, which is representative of the whole system;  {\bf field} $(pq)$-particles  in  $\bx^*$ of the visibility zone $\Lambda$; and  {\bf candidate} pedestrian $(hk)$, in  $\bx$.
The candidate pedestrian modifies, in probability, its state into that of the test pedestrian, due to interactions with the field pedestrians, while the test pedestrian losses its state due to interactions with the field pedestrians.
The activity of the candidate pedestrian is $u_w$, which is not modified by the interaction.

\sm \no iii) The output of interactions is modeled, by using tools of the game theory \cite{[SPL06],[SVS12]}, by the discrete probability density function $\cA_{hk,pq}^\si (ij)$, which denotes the probability  that a candidate $(hk)$-particle modifies its state into that of the test $(ij)$-particle due to interaction with the field $(pq)$-particles. The transition density depends on the whole local number density $\rho$.

\sm \no iv) The frequency of the interactions is modeled by the term $\eta$ that is allowed to depend on the density of the field particles in the visibility zone.

\sm

Time and space dynamics of the distribution function $f_{ij}^\sigma$ can be obtained by equating its increase on time in the elementary volume of the space of the microscopic states to the net flow into such volume due to transport and interactions:

\begin{eqnarray}
 \big(\partial_{t} &+& \bv_{ij} \cdot
 \partial_{\bf x} \big) f_{ij}^\si(t,\bx) = \cJ [\f](t,\bx)\hfill \label{eq_gral}\\
&=&  \sum_{h,p=1}^n   \sum_{k,q=1}^m \int_\Lambda \eta[\rho(t,\bx^*)]
  \cA_{hk,pq}^\si (ij)[\rho(t,\bx^*)]  f_{hk}^\si(t,\bx) \, f_{pq}^\si(t, \bx^*)\,d\bx^*
  \nonumber\\0000
&-&  f_{ij}^\si(t,\bx)\sum_{p=1}^n   \sum_{q=1}^m  \int_\Lambda
\eta[\rho(t,{\bx}^*)] \, f_{pq}^\si(t,\bx^*)\,d\bx^*, \label{SM1}
\end{eqnarray}
where $\f = \{f_{ij} \}$, while the term $\cA_{hk,pq}^\si (ij)$
should be consistent with the probability density property:
\begin{equation} \label{pr}
\sum_{i=1}^n \sum_{j=1}^m \cA_{hk,pq}^\si (ij) = 1, \quad \forall
\, hp \in \{1, \ldots, n\}, \quad \forall \, kq \in \{1, \ldots,
m\} \vb
\end{equation}
for $\si = 1,2$, and for all conditioning local density. The derivation of specific models can take advantage of the above structures and is obtained by modeling the various terms that appear in Eq.~(\ref{SM1}). The modeling necessarily needs heuristic assumptions towards the interpretation of the complex
phenomenology of the system under consideration.

\sm \no $\bullet$ {\bf Interaction rate:} The modeling of the term $\eta$, can be developed similarly to the case of vehicular traffic \cite{[BDF12]},  by increasing the interaction rate with increasing local density in the free and congested regimes. For higher densities, when  pedestrians are obliged to stop, one may assume that $\eta$ keeps a constant value, or may decay for lack of interest. The following model can be proposed among various conceivable ones:
\begin{equation}\label{rho}
\eta (\rho(t,\bx)) = \eta^0 (1 + \rho(t,\bx)) \exp\big(- \rho(t,\bx)\big)\vb
\end{equation}
while a simpler model can be obtained by assuming  $\eta = \eta^0$.

\sm \no  $\bullet$ {\bf Transition probability density:} The modeling of the  transition probability density $\cA_{hk,pq}^\si (ij)$ refers to the interactions involving candidate and field particles. The approach proposed here is based on the assumption that particles are subject to three different influences, namely the {\sl trend to the exit point}, the {\sl influence of the stream}  induced by the other pedestrians, and the selection of the path with minimal density gradient.  A simplified interpretation of the phenomenological behavior is obtained by assuming the factorization of the two probability densities modeling the modifications of the velocity direction and modulus:

\begin{equation}\label{factorization}
 \cA_{hk,pq}^\si (ij) =\cB_{hp}^\si (i) \big(\th_h \to \th_i|\rho(t,\bx) \big)\times \cC_{kq}^\si (j) \big(v_k \to v_j|\rho(t,\bx) \big),
\end{equation}
where subscripts correspond to the state of the interacting pairs, superscripts to the group and the output of the interaction is indicated in the argument, which depends on the local density.

\sm \no {\bf Remark 2.3.}
 Equation (\ref{factorization}) is based on an heuristic assumption that cannot be justified by theoretical issues. However, it simplifies the modeling approach by separating the various causes that modify the dynamics.  This assumption does not imply factorization of the probability density due to the mixing action of the variable $\rho$.

Bearing this remark in mind, let us consider, separately the modeling of the two probability densities:

\sm \no -- {\bf Modeling} $\cB^\si_{hp}(i):$ Let us consider the candidate particle in the position $P$ moving in the direction $\th_h$ and interacting with a field particle with direction $\th_p$. Moreover,  let $\th_\nu$ be the angle from $P$ to $T_{\si}$  or to the direction of the shortest path, see Figure \ref{fig: visibility_interactions}(b). Let us introduce a parameter $\a_\si$, which models the sensitivity of pedestrians both to reach the target $T_\si$, and to follow the stream.  Moreover,  it is assumed that the  sensitivity to the target, depends on the ratio of vacuum $1 - \rho$, while the sensitivity to the stream depends on the density $\rho$. Finally it is assumed that the dynamics is more active for higher values of the activity variable $u_0$. It is natural that an active individual shows a higher ability to modify the direction of the motion. The games modeling interactions are the same for both groups being understood that the target chased by pedestrians differs for each of them.  This qualitative description can be formalized as follows:

\sm \no -- {\sl Interaction with a upper stream and target  directions, namely $\th_p > \th_h, \quad \th_\nu > \th_h$:} It is assumed that, in
 addition to the selection of the path with minimal gradients, both actions contribute to an anticlockwise rotation:
\begin{eqnarray*}
\cB^\si_{hp} (i) & = & \a_\si \, u_0(1-\rho) + \a_\si \, u_0\,\rho \quad  \hbox{if} \quad   i = h + 1\vb  \\
\cB^\si_{hp} (i)  & = & 1 - \a_\si \, u_0(1-\rho)  -   \a_\si \, u_0 \,\rho \quad  \hbox{if} \quad  i = h \vb  \\
\cB^\si_{hp} (i) & = & 0  \quad  \hbox{if} \quad   i = h - 1 \pb
\end{eqnarray*}

\sm Analogous calculations can be done for the other cases. The result is as follows:

\sm \no -- {\sl Interaction with a upper stream and low target direction $\th_p > \th_h; \quad \th_\nu < \th_h$:}
\begin{eqnarray*}
\cB_{hp}^\si (i)  & = &    \a_\si \, u_0\,\rho  \quad  \hbox{if} \quad   i = h + 1\vb  \\
\cB_{hp}^\si (i)   & = &  1 -  \a_\si \, u_0(1-\rho) -   \a_\si \, u_0 \,\rho \quad  \hbox{if} \quad  i = h \vb  \\
\cB_{hp}^\si  (i)  & = &  \a_\si \, u_0\, (1-\rho)  \quad  \hbox{if} \quad  i = h - 1\pb
\end{eqnarray*}

\sm \no -- {\sl Interaction with a lower stream and upper target direction $\th_p < \th_h; \quad \th_\nu > \th_h$:}
\begin{eqnarray*}
\cB_{hp}^\si (i)   & = & \a_\si \, u_0 (1-\rho)  \quad  \hbox{if} \quad   i = h + 1\vb  \\
\cB_{hp}^\si (i)   & = &  1 - \a_\si \, u_0(1-\rho)  - \a_\si \, u_0 \,  \rho \quad  \hbox{if} \quad  i = h \vb  \\
\cB_{hp}^\si  (i)  & = &   \a_\si \, u_0\, \rho \quad  \hbox{if} \quad  i = h - 1\pb  \\
\end{eqnarray*}

\sm \no -- {\sl Interaction with a lower stream and target directions $\th_p < \th_h; \quad \th_\nu < \th_h$:}
\begin{eqnarray*}
\cB_{hp}^\si  (i)  & = & 0  \quad  \hbox{if} \quad   i = h + 1\vb  \\
\cB_{hp}^\si (i)   & = & 1 -  \a_\si \, u_0(1-\rho) -  \a_\si \, u_0 \, \rho \quad  \hbox{if} \quad    i = h \vb  \\
\cB_{hp}^\si (i)   & = &  \a_\si \, u_0\, (1-\rho)   + \a_\si \, u_0\,  \rho \quad  \hbox{if} \quad   i = h - 1 \pb  \\
\end{eqnarray*}

\sm

Concerning the term $\cC_{kq}^{\si}(j)$, the proposed modeling assumes that pedestrians have a trend to adjust each other their velocity modulus,
  namely an encounter with a faster (slower) pedestrian induces a trend to increase (decrease) the speed. This trend is conditioned by the local density, namely it decreases with increasing density. It is assumed that also this type of dynamics depends on the activity $u_0$ of the pedestrian and on their sensitivity $\b_\si$ to the change of velocity. The corresponding table of games, which defines the interaction rules, is as follows:

\sm \no $\bullet$ \textit{ Interaction with faster (or slower) particles $v_k < v_q$, or  $v_k > v_q$, respectively}
 $$
\cC_{kq}^{\si}(j)=\left\{
           \begin{array}{ll}
             1-\b_\si\, u_0 \rho, \quad  & \hbox{$j = k$;} \\
              \b_\si u_0 \rho, \quad  & \hbox{$j={k+1}$;} \\
             0, \quad  & \hbox{otherwise;}
           \end{array}\nonumber
         \right.
$$

$$
\cC_{kq}^{\si}(j)=\left\{
           \begin {array}{ll}
             \b_\si \, u_0 \rho, \quad  & \hbox{$j = k$;} \\
              1-\b_\si\,  u_0 \rho, \quad  & \hbox{$j ={k-1}$;} \\
             0, \quad  & \hbox{otherwise.}
           \end{array}
         \right.
         $$

\sm \no   $\bullet$ \textit{Interaction with equally fast  particles $k = q$}

$$
\cC_{kq}^{\si}(j)=\left\{
           \begin{array}{ll}
             1- 2 \b_\si \,u_0 \rho, \quad  & \hbox{$j=k$;} \\
               \b_\si \,u_0 \rho, \quad  & \hbox{$j=k - 1$;} \\
              \b_\si \,u_0 \rho, \quad  & \hbox{$j= k + 1$.}
           \end{array}
         \right.\nonumber
$$

Moreover,

\sm \no$\bullet$  \textit{for  $k=1$ the candidate particle cannot reduce the velocity, while for  $k=k$
cannot increase it}:
$$
\cC_{kq}^{\si}(j) =\left\{
           \begin{array}{ll}
             1-\b_\si \,u_0\,\rho, \quad  & \hbox{$j=1$;} \\
              \b_\si \,u_0\,\rho, \quad  & \hbox{$j=2$;} \\
             0,  \quad  & \hbox{otherwise;}
           \end{array}
         \right. \nonumber
$$

$$
 \cC_{kq}^{\si}(j) =\left\{
           \begin{array}{ll}
             \b_\si \,u_0\,\rho, \quad  & \hbox{$j=m -1$;} \\
              1-\b_\si \,u_0\,\rho, \quad  & \hbox{$j=m$;} \\
             0 , \quad  & \hbox{otherwise.}
           \end{array}\nonumber
         \right.\nonumber
$$

\sm \no$\bullet$  \textit{The modeling corresponding to  jam densities  $\rho \in [\ve,1]$  implements that pedestrians reduce to zero their velocity.}

\sm \no {\bf Remark 2.4.} If the heterogeneous behavior of pedestrians needs to be taken into account a probability distribution over the variable $u$
has to be introduced with the modeling of the interaction dynamics that modifies it.

\vskip.5truecm


\section{On the Initial Value Problem}

The initial value problem consists in solving Eqs.~(\ref{SM1}) with initial conditions given by
\begin{equation} \label{ini}
f_{ij}^\si(0,\bx)=\phi_{ij}^\si(\bx).
\end{equation}

Let us introduce the mild form obtained by integrating along the
characteristics:
\begin{eqnarray}\label{mild}
\widehat{f_{ij}^\si}(t,\bx)&=& \phi_{ij}^\si(\bx) + \int_0^t
\bigg(\widehat{\Gamma_{ij}^\si}[\f,\f](s,\bx)  -
\widehat{f_{ij}^\si}(s,\,\bx) \widehat{L[\f]}(s,\,\bx)\bigg)ds,
\nonumber \\ & & \quad i \in \{1, \ldots, n\}, \quad  \, j \in
\{1, \ldots, m\}, \quad \sigma \in \{1,2\},
\end{eqnarray}
where the following notation has been used for any given vector $f(t, \bx)$:
$\widehat{f_{ij}^\si}(t,\bx)=f_{ij}^\si(t,x+v_j \cos(\th_i) t,
y+v_j \sin(\th_i)t)$, and where the operators in (\ref{mild}) are defined as follows:

\begin{eqnarray}\label{gain} 
& \widehat{\Gamma_{ij}^\si[\f,\f]}&(t,\bx) =  \sum_{h,p=1}^n   \sum_{k,q=1}^m
\int_{\Lambda} \eta[\rho(t,\bx^*)]
  \cA_{hk,pq}^{\si} (ij)[\rho(t,\bx^*)] \nonumber \\  &   \widehat{f_{hk}^{\si}}&(t,x+(v_j
   \cos(\th_i) -v_k \cos(\th_h))t,y+(v_j \sin(\th_i) -v_k \sin(\th_h))t ) \,
  \nonumber \\ & \widehat{f_{pq}^{\si}}&(t, x^*-v_q \cos(\th_p)t,y^*-v_q
  \sin(\th_p)t
  )\,d\bx^*,\end{eqnarray}
  and
 \begin{equation} \label{loss}\widehat{L [\f]}(t,\,\bx)=\sum_{p=1}^n   \sum_{q=1}^m  \int_{\Lambda} \eta[\rho(t,{\bx}^*)] \,
\widehat{f_{pq}^\si}(t,x^*-v_q \cos(\th_p)t,y^*-v_q \sin(\th_p)t
)\,d\bx^*,
\end{equation}
where $\f = \{f_{ij}^\si \}$.

Let us now   define  the following functional space:
$$ L^1_{M_{2n,2m}}= \{ f=(f_{ij}^\si)\in
M_{2n, 2m} :
\parallel f
\parallel_1= \sum_{\sigma=1}^{2}
 \sum_{i=1}^{n} \sum_{j=1}^m \int_{\Omega}
 \mid f_{ij}^\si(t, \bx) \mid \, d\bx < \infty
\}$$
and introduce, for $T>0$, the Banach space
$$X_T=X[0,T]=C([0,T],L^1_{M_{2n,2m}})
$$
of the matrix-valued functions $f=f(t,\bx): [0,T] \times \Omega \rightarrow M_{2n,2m}.$
 such that for all fixed $t\in [0,T]$ the function $\bx\rightarrow f(t,\bx)$ belongs to
$L^1_{M_{2n,2m}}$, where the norm in $X_T$ is given by
 \begin{equation}\label{norm}
\parallel f \parallel_{X_T}=\sup_{t\in[0,\,T]} \parallel f \parallel_1.
 \end{equation}

The qualitative analysis developed in the following takes
advantage of the following assumptions that are consistent with the physics of the system under consideration:

\sm \noindent{\bf Assumption H.1.} For all positive $R$, there
exists a constant $C_{\eta}>0$ so that
\begin{equation} 0< \eta(\rho)\leq C_{\eta}\vb \quad \hbox{whenever} \quad  0\leq \rho\leq R\pb
 \end{equation}

\sm \noindent{\bf Assumption H.2.} Both the encounter rate
$\eta[\rho]$ and the transition probability $\cA_{hk,pq}^\si
(ij)[\rho]$ are Lipschitz continuous functions of the macroscopic
density $\rho$, i.e., that there exist constants $L_{\eta}, L_{
\mathcal{A}}$ such that
\begin{equation} \left\{
\begin{array}{l}
\displaystyle{ \mid  \eta[\rho_1]- \eta[\rho_2]\mid \leq L_{\eta}
\mid \rho_1 - \rho_2 \mid},\\
   {}  \nonumber  \\
\displaystyle{
 \mid \cA_{hk,pq}^\si (ij)[\rho_1]- \cA_{hk,pq}^\si (ij)[\rho_2] \mid \leq L_{
\mathcal{A}} \mid \rho_1 - \rho_2 \mid}, \nonumber
\end{array} \right.
\end{equation} whenever
$0\leq \rho_1\leq R$, $0\leq \rho_2\leq R$, and all $i, h,p
=1,..,n$ and $j, k, q=1,..,m$.\vskip0,1cm

The following results can be proved under  assumptions H..1 and H.2:

\sm \begin{theorem} ({\bf Local Existence}) Let $\phi_{ij}^\si \in
L_\infty \cap L^1$, $\phi_{ij}^\si \geq 0$, then there exists
$\phi^0$ so that, if $\parallel \phi\parallel_1 \leq \phi^0 $,
there exist $T$, $a_0$, and $R$ so that a unique non-negative
solution to the initial value problem of
Eqs.(\ref{SM1})--(\ref{ini}) exists and satisfies:
\begin{equation}f \in X_T, \qquad
 \sup_{t\in [0,T]}  \parallel  f (t)
\parallel_1 \leq a_0 \parallel \phi \parallel_1,
\end{equation}
 and
 \begin{equation} \label{bou} \rho(t,\bx) \leq
R, \quad \forall t\in [0,T], \quad \bx \in\Omega  \pb
\end{equation}
Moreover, if
\begin{equation}\label{bound2}
\sum_{\sigma=1}^2 \sum_{i=1}^n \sum_{ j=1}^m
\parallel \phi_{ij}^\si \parallel_\infty \leq 1\vb
\end{equation} and $\parallel \phi \parallel_1$ is small,
one has
\begin{equation}\label{bound3}
 \rho(t,\bx) \leq 1,  \quad \forall t\in [0,T], \quad \bx \in\Omega .
 \end{equation}

\end{theorem}

\sm \begin{theorem} ({\bf Global Existence}) Under the assumptions
of Theorem 3.1,   there exist $\phi^r$, $(r=1,...,p-1)$ such that
if $\parallel \phi
\parallel_1 \leq \phi^r$, there exists $a_r$ so that it is possible to find a unique non-negative solution to the
initial value problem of Eqs.(\ref{SM1})--(\ref{ini}) satisfying
for any $r \leq p-1$
\begin{equation}\label{kk}
f(t) \in X[0,(p-1)T],
\end{equation}
\begin{equation}\label{zz}
\sup_{t\in [0,T]}  \parallel  f (t+(r-1)T)
\parallel_1 \leq a_{r-1}
\parallel \phi
\parallel_1,
\end{equation}
and
 \begin{equation} \label{bound4}
 \rho(t+(r-1)T,\bx) \leq R,  \quad \forall t\in [0,T], \quad \bx \in\Omega .
 \end{equation}
\sm \noindent Moreover, if $\phi$ satisfies (\ref{bound2}), one
has

\begin{equation}\label{bound5}
 \rho(t+(r-1)T,\bx) \leq 1,  \quad \forall t\in [0,T], \quad \bx \in\Omega .
 \end{equation}

 \end{theorem}

 \vskip0.5cm


\section{Proof of Theorem 3.1 (Local Existence)}

 Let us introduce the function
\begin{equation}
\psi_{ij}^\si(t,\bx)= \exp(\lambda t ) f_{ij}^\si(t,\bx)\vb \quad
\hbox{for} \quad \lambda >0\pb
\end{equation}
Therefore, system (\ref{SM1})-(\ref{ini}) can be written in
equivalent form in terms of the functions
$\psi(t,x)=(\psi_{ij}^\si(t,x))$, where $\psi_{ij}^\si$ are
solutions of the following mild problem:
\begin{eqnarray}\label{mivp}
 \widehat{\psi_{ij}^\si}(t,\bx) &=& \phi_{ij}^\si(\bx) + \int_0^t \bigg(\widehat{\Gamma_{ij}^\si[\psi,\psi]}(s,\bx) \exp(-\lambda s)
  \nonumber \\ & + & \displaystyle \widehat{\psi_{ij}^\si}(s,\,\bx)\bigg\{\lambda - \widehat{L
[\psi]}(s,\,\bx)\exp(-\lambda s)\bigg\}\bigg)ds.
\end{eqnarray}

Moreover, let us consider the operator  $A$  acting on $X_T$ whose components are
\begin{eqnarray}\label{A}
\widehat{A(\psi)_{ij}^\si}(t,\bx)&=& \phi_{ij}^\si(\bx) + \int_0^t
\bigg(\widehat{\Gamma_{ij}^\si[\psi,\psi]}(s,\bx)\exp(-\lambda s)
\nonumber \\ &+&\displaystyle
\widehat{\psi_{ij}^\si}(s,\,\bx)\bigg\{\lambda - \widehat{L
[\psi]}(s,\,\bx)\exp(-\lambda s)\bigg\}\bigg)ds.
\end{eqnarray}

The principle of contractive mapping can be applied to
Eq.(\ref{A}). However, the proof is based on
various sharp inequalities that are preliminary proved in the
following Lemmas.

\me\begin{lemma} Let $T>0$, $\psi\in X_T$, then $ A\psi \in X_T$
and $\exists C_1>0$ so that
\begin{equation}\label{esti1}
\parallel A(\psi) \parallel_{X_T}\leq \parallel
\phi \parallel_1 +{C_1 \over \lambda } \parallel \psi
\parallel_{X_T}^2 +\lambda T \parallel \psi
\parallel_{X_T}.
\end{equation}
\end{lemma}

\sm \noindent {\bf Proof of Lemma 4.1:}  To prove the estimate
(\ref{esti1}), we need first to estimate the nonlinear operators
$\widehat{\Gamma_{ij}^\si}[\psi,\psi]$ and $\widehat{ L [\psi]}$
in $L_1$. Taking into account the assumption H.1 on $\eta$ yields
\begin{eqnarray}
 \parallel\widehat{\Gamma_{ij}^\si}[\psi,\psi] \parallel_1 &\leq &  C_\eta \sum_{h,p=1}^n   \sum_{k,q=1}^m
\int_{\Lambda} \cA_{hk,pq}^\si (ij)[\rho(t,\bx^*] \nonumber \\ &\times&
 \mid \widehat{\psi_{pq}^\si}(t, x^*-v_q
\cos(\th_p)t,y^*-v_q \sin(\th_p)t)\mid \,d\bx^* \nonumber \\
&\times& \int_\Omega \mid \widehat{\psi_{hk}^\si}(t,x+(v_j \cos(\th_i) -v_k \cos(\th_h))t, y \nonumber \\
 &+&(v_j \sin(\th_i) -v_k \sin(\th_h))t )\mid \, d\bx.
   \end{eqnarray}

 Using Eq. (\ref{pr})  on   the probability density
$\cA_{hk,pq}^\si (ij)$
 and the following transformations with jacobian equal one
$$ \bx=(x,y)\mapsto (x-v_q \cos(\th_p)t,y-v_q
  \sin(\th_p)t), $$
$$\bx= (x,y)\mapsto (x+(v_j \cos(\th_i) -v_k \cos(\th_h))t,y+(v_j
\sin(\th_i) -v_k \sin(\th_h))t ), $$ yields:
\begin{equation} \label{ga} \parallel \widehat{\Gamma}[\psi,\psi] \parallel_1
=\sum_{\sigma=1}^2 \sum_{i,j} \parallel
\widehat{\Gamma_{ij}^\si}[\psi,\psi] \parallel_1 \leq C_\eta
\parallel \psi \parallel_1^2. \end{equation}

\noindent Moreover, the same estimate  can be obtained  for the
loss operator $\widehat{L}$:
  \begin{equation} \label{lo}
  \parallel \widehat{ L [\psi]}\widehat{\psi} \parallel_1
 \leq C_\eta \parallel \psi \parallel_1^2\pb
 \end{equation}

 The following inequality is directly deduced, by  using
(\ref{ga})-(\ref{lo}), from (\ref{A}):
\begin{equation}\label{C1}
\parallel A(\psi) \parallel_1 \leq \parallel \phi \parallel_1 +
{2  \over \lambda} C_\eta (1- \exp (-\lambda t)) \parallel \psi
\parallel_{X_T}^2 + \lambda \, t \parallel \psi
\parallel_{X_T},\nonumber
\end{equation}
which gives (\ref{esti1}) with $C_1=2\,C_\eta.$

Analogous calculations lead to the following inequality:
$$
\parallel A(\psi)(t)- A(\psi)(s)\parallel_1 \leq \mid t-s\mid (  C_1
\parallel \psi \parallel_{X_T}^2 +\lambda  \parallel \psi \parallel_{X_T})\vb
$$
which gives the continuity of the operator $A$ in $L_1$.
Therefore, Lemma 4.1 is proved. $\, \bull$

\me

\begin{lemma}  Let  $\psi^1=({\psi_{ij}^\si}^1)\in L_1, \psi^2=({\psi_{ij}^\si}^2) \in L_1$, such that
\begin{equation} \label{est}
\left\{
\begin{array}{l}
\displaystyle{\sum_{\sigma=1}^2
\sum_{i=1}^{n}\sum_{j=1}^{m}}\widehat{{\psi_{ij}^\si}^1}(t,\,
x-v_j \cos(\th_i) t, y-v_j \sin(\th_i) t)\leq R \exp(\lambda t),\\
   {}  \nonumber  \\
\displaystyle{\sum_{\sigma=1}^2
\sum_{i=1}^{n}\sum_{j=1}^{m}}\widehat{{\psi_{ij}^\si}^2}(t,\, x-v_j
\cos(\th_i) t, y-v_j \sin(\th_i) t)\leq R \exp(\lambda t), \nonumber
\end{array} \right.
\end{equation}
for $x \in \Omega, t \geq 0$. Then
\begin{eqnarray}
\label{ga1}
\parallel \widehat{\Gamma}(\psi^1, \psi^1)-
\widehat{\Gamma}(\psi^2, \psi^2)\parallel_1 & \leq & (C_{\eta} +
nm R(C_\eta
L_\mathcal{A} +L_\eta)) \nonumber\\
& \times & (\parallel \psi^1 \parallel_{1} + \parallel \psi^2
\parallel_{1})
\parallel \psi^1-\psi^2\parallel_{1} \vb
\end{eqnarray}
\begin{equation}
\label{ga2} \parallel \widehat {\psi^1}\,
\widehat{\Lambda}(\psi^1)-\widehat {\psi^2}\,
\widehat{\Lambda}(\psi^2)\parallel_1 \leq
 (C_\eta +L_\eta R  )
(\parallel \psi^1\parallel_{1}+\parallel \psi^2\parallel_{1})
\parallel \psi^1-\psi^2
\parallel_{1}.
\end{equation}
\end{lemma}

\sm {\bf Proof of Lemma 4.2:}  First write $\widehat{\Gamma_{ij}^\si}(\psi^1, \psi^1)(t,\bx)-
\widehat{\Gamma_{ij}^\si}(\psi^2,\psi^2)(t,\bx)$ as follows:
\begin{eqnarray*}
&{}&  \widehat{\Gamma_{ij}^\si}(\psi^1, \psi^1)(t,\bx) - \widehat{\Gamma_{ij}^\si}(\psi^2,\psi^2)(t,\bx)  \nonumber\\
 &{}& \quad  = \sum_{h, p=1}^n \sum_{k,
q=1}^m\int_{\Lambda}\eta[\rho_{\psi^1}(t,\bx^*)]\cA_{hk,pq}^{\si}(ij)[\rho_{\psi^1}(t,\bx^*)] \nonumber \\
 &{}&   \quad \times \bigg[ \widehat{{\psi_{hk}^\si}^1}(t,x+(v_j \cos(\th_i) -v_k \cos(\th_h))t,y+(v_j \sin(\th_i) -v_k
\sin(\th_h))t ) \, \nonumber
\\&&  \quad  \times \widehat{{\psi_{pq}^\si}^1}(t, x^*-v_q \cos(\th_p)t,y^*-v_q
  \sin(\th_p)t)\,\nonumber \\
&{}& \quad   -  \widehat{{\psi_{hk}^\si}^2}(t,x+(v_j \cos(\th_i) -v_k \cos(\th_h))t,y+(v_j \sin(\th_i) -v_k
\sin(\th_h))t ) \,
\nonumber\\
&& \quad  \times \widehat{{\psi_{pq}^\si}^2}(t, x^*-v_q \cos(\th_p)t,y^*-v_q
  \sin(\th_p)t)\,\bigg]\,d\bx^*  \nonumber \\
&{}& \quad +   \sum_{h, p=1}^n \sum_{k, q=1}^m\int_{\Lambda} \widehat{{\psi_{h,k}^\si}^2}(t,x+(v_j \cos(\th_i) -v_k
\cos(\th_h))t,y \nonumber \\
&& \quad  +(v_j \sin(\th_i) -v_k \sin(\th_h))t )\nonumber \\ && \quad \times \widehat{{\psi_{p,q}^\si}^2}(t, x^*-v_q
\cos(\th_p)t,y^*-v_q
  \sin(\th_p)t
  )\nonumber \\
&{}&  \quad \times \bigg[\eta[\rho_{\psi^1}(t,x^*)]
 \cA_{hk,pq}^{\si}(ij)[\rho_{\psi^1}(t,x^*)] \nonumber \\
&{}& \quad - \eta[\rho_{\psi^2}(t,x^* )]
 \cA_{hk,pq}^{\si}(ij)[\rho_{\psi^2}(t,\,x^*)]\bigg]dx^* \nonumber \\
&{}& \qquad =  A_{ij}^\si +B_{ij}^\si, \nonumber
\end{eqnarray*}

Using Assumptions H.1, H.2,  and summing over $i,j, \sigma$, yields
\begin{eqnarray}
\label{b1}  & & \qquad \sum_{\sigma=1}^2 \sum_{i=1}^n  \sum_{j=1}^m
\parallel A_{ij}^\si
\parallel_1 \nonumber \\
 && \quad \leq   C_\eta \sum_{\sigma=1}^2
   \sum_{i=1}^n \sum_{j=1}^m\sum_{h, p=1}^n \sum_{k, q=1}^m
\int_{\Lambda\times \Omega} \cA_{hk,pq}^{\si}(ij) \mid \widehat{{\psi_{h,k}^\si}^1}(t,x+(v_j \cos(\th_i) \nonumber \\
&& \quad -v_k \cos(\th_h))t,  y+(v_j \sin(\th_i) -v_k \sin(\th_h))t )\mid \nonumber\\   && \quad  \times  \mid
\widehat{{\psi_{pq}^\si}^1}(t, x^*-v_q \cos(\th_p)t,y^*-v_q
  \sin(\th_p)t
  )  \nonumber \\&&  \quad -  \widehat{{\psi_{pq}^\si}^2}(t, x^*-v_q \cos(\th_p)t,y^*-v_q
  \sin(\th_p)t)\mid \,d\bx^* d\bx \nonumber\\   &&  \quad + C_\eta \sum_{\sigma=1}^2 \sum_{i=1}^n
\sum_{j=1}^m\sum_{h, p=1}^n \sum_{k, q=1}^m \int_{\Lambda \times \Omega} \cA_{hk,pq}^{\si}(ij) \nonumber \\
&&\quad\times \mid
\widehat{{\psi_{pq}^\si}^2}(t, x^* -v_q \cos(\th_p)t,y^*- v_q
  \sin(\th_p)t
  )\mid \nonumber \\   &&  \quad  \times
 \mid \widehat{{\psi_{h,k}^\si}^1}(t,x+(v_j \cos(\th_i) -v_k
\cos(\th_h))t,y \nonumber \\
&&\quad +(v_j \sin(\th_i) -v_k \sin(\th_h))t )\nonumber
\\ &&  \quad -  \widehat{{\psi_{h,k}^\si}^2}(t,x+(v_j \cos(\th_i) -v_k
\cos(\th_h))t,y \nonumber
\\ &&  \quad+(v_j \sin(\th_i) -v_k \sin(\th_h))t )\mid  \,d\bx^* d\bx  \nonumber \\
&&  \qquad   \leq C_\eta
\parallel \psi^1-\psi^2\parallel_1(
\parallel \psi^1\parallel_1+\parallel \psi^2\parallel_1) \pb
\end{eqnarray}
On the other hand, considering that
 \begin{eqnarray*}
&&\mid (\rho_{\psi^1} -\rho_{\psi^2})(t,\bx^*) \mid  \leq 
\exp(-\lambda t)\sum_{\sigma=1}^2 \sum_{r=1}^n \sum_{\nu=1}^m \mid
\widehat{{\psi_{r \nu}^\si}^1}(t,\, x^*-v_\nu \cos(\th_r) t,
y^* \nonumber
\\ &&  \quad -v_\nu \sin(\th_r) t) \widehat{{\psi_{r \nu}^\si}^2}(t,\, x^*-v_\nu \cos(\th_r) t, y^*-v_\nu \sin(\th_r)
t)\mid \vb
\end{eqnarray*}
and
 \begin{equation}
 \sum_{p=1}^n
\sum_{q=1}^m \mid\widehat{{\psi_{pq}^\si}^2}(t, x^*-v_q
\cos(\th_p)t,y^*-v_q
  \sin(\th_p)t) \mid \leq
R \exp(\lambda t), \nonumber
\end{equation}

then, using Assumptions H.1-H.2, noting that
$\cA_{hk,pq}^{\si}(ij) \leq 1$, and summing over $i,j, \sigma$,
yields:
\begin{eqnarray*} \sum_{\sigma=1}^2 \sum_{i=1}^n
\sum_{j=1}^m \parallel B_{ij}^\si \parallel_1 \leq \leq  nm
R(L_\eta +C_\eta L_A)
\parallel \psi^2
\parallel_{1} \parallel \psi^1-\psi^2
\parallel_{1}\end{eqnarray*}
which gives, by using (\ref{b1}), the estimate (\ref{ga1}).

By the same arguments, from Assumptions H.1, and  H.2 one deduces
(\ref{ga2}) and this completes the proof of Lemma 4.2.  $\, \bull$

\me
\begin{lemma} Let $T>0$, $\psi^1= ({\psi_{ij}^\si}^1),
\psi^2=({\psi_{ij}^\si}^2) \in X_T$, satisfying (\ref{est}) for $x
\in\Omega,t \in [0,T] $. Then:

\sm\noindent  1. There exists a constant  $C_2>0$ such that
\begin{eqnarray} \label{ab}  \parallel A(\psi^1)-
A(\psi^2)\parallel_{X_T} &\leq & \bigg({C_2 \over \lambda
}(\parallel \psi^1
\parallel_{X_T}+ \parallel \psi^2
\parallel_{X_T}) +\lambda T \bigg)  \nonumber \\ & &  \parallel \psi^1-\psi^2
\parallel_{X_T} \pb \end{eqnarray}

\sm\noindent 2. If ${(\psi^1)}_{ij}^{\si} \geq 0$, then there
$\exists \lambda_0$ such that if $\lambda \geq \lambda_0$ and if
$\phi_{ij}^\si\geq 0$ one has $\widehat{(A(\psi^1))_{ij}^\si}\geq
0$.

\sm \noindent 3. Let $\phi_{ij}^\si \in L^\infty$, then for  $R $ large enough($ R \geq R_{1}$),  there
exists $T$ such that
\begin{equation} \label{es2} \sum_{\sigma=1}^2 \sum_{i=1}^n \sum_{j=1}^m \widehat{(A(\psi^1))_{ij}^\si}(t,\, x-v_j
\cos(\th_i) t, y-v_j \sin(\th_i) t)\leq R \exp(\lambda
t),\end{equation}

\noindent $ \forall t\in [0,T], \quad \forall x\in \Omega. $

\end{lemma}

\sm \noindent{\bf Proof of Lemma 4.3:} From (\ref{A}) and  Lemma 4.1,  one
gets easily (\ref{ab}) with
 \begin{equation} \label{C2} C_2=
C_{\eta} + n\, m \, R(C_\eta L_\mathcal{A} +L_\eta).
\end{equation}
 Since
$\widehat{\Gamma_{ij}^\si}[\psi^1,\psi^1](t,\bx)\geq
 0$ if $({\psi^1})_{i,j}^\si\geq 0$,  the nonnegativity  of
 $\widehat{(A(\psi^1))_{ij}^\si}$ depend on the possibility to find $\lambda
 >0$ such that
\begin{equation}
\lambda - \widehat{ L [\psi^1]}(s,\,\bx) \exp(-\lambda s)\geq 0.
\end{equation}

Noting that from (\ref{loss}) and  assumption H.1, one
obtains
\begin{eqnarray*} 
&&\widehat{L
[\psi^1]}(t,\,x) = \sum_{p=1}^n   \sum_{q=1}^m  \int_\Lambda
\eta[\rho(t,{\bx}^*)] \, \widehat{{\psi^1}_{pq}^\si}(t,x^{*} - v_q
\cos(\th_p)t,y^{*} \nonumber
\\ &&  \quad -v_q \sin(\th_p)t )\,d\bx^* \leq  R  \exp(\lambda t) C_\eta  \mid \Lambda \mid. \nonumber
\end{eqnarray*}

The nonnegativity of $\widehat{(A(\psi^1))_{ij}^\si}$ is
then achieved by choosing $\lambda \geq \lambda_0= R C_\eta \mid
\Lambda \mid $. \noindent To deal with 3) first let see that
\begin{eqnarray}
&& \widehat{\Gamma}(\psi^1, \psi^1)(t,\, x-v_j \cos(\th_i) t, y-v_j \sin(\th_i) t)\nonumber \\
&& \qquad  = \sum_{h, p=1}^n \sum_{k,q=1}^m\int_{\Lambda} \eta[\rho(t,x^*)]
\cA_{hk,pq}^{\si}(ij)[\rho(t,x^*)]\nonumber \\
 && \qquad  \times\widehat{{\psi^1}_{h,k}^\si}(t,x -v_k \cos(\th_h)t,y
 -v_k \sin(\th_h)t )\nonumber \\
 && \qquad \times \widehat{{\psi^1}_{p,q}^\si}(t, x^*-v_q \cos(\th_p)t,y^*-v_q \sin(\th_p)t)d\bx^*\nonumber \\
&& \qquad \leq C_\eta R^2 \exp(2\lambda t) \int_{\Lambda}d\bx^* \vb  \leq   C_\eta \mid \Lambda \mid  R^2 \exp(2\lambda t).
\end{eqnarray}

Then  the following estimate is proved:
\begin{eqnarray*} && \sum_{\sigma=1}^2 \sum_{i=1}^n
\sum_{ j=1}^m \widehat{(A(\psi))_{ij}^\si}(t,\, x-v_j \cos(\th_i)
t, y-v_j \sin(\th_i) t)\nonumber \\ && \quad \leq  \sum_{\sigma=1}^2 \sum_{i=1}^n \sum_{
j=1}^m \parallel \phi_{ij}^\si\parallel_\infty  +{1 \over 2 \lambda} n\, m\,
C_\eta \mid \Lambda \mid  R^2 (\exp(2\lambda t)-1) \nonumber
\\ &&  \quad + R(\exp(\lambda t)-1)\vb
\end{eqnarray*}
which gives (\ref{es2}) if
 \begin{equation}\label{cc}
 R > \sum_{\sigma=1}^2  \sum_{i=1}^n \sum_{j=1}^m
\parallel \phi_{ij}^\si \parallel_\infty = R_1,
\end{equation}
and
\begin{equation}\label{cc1}  t\leq T={1 \over {2 \lambda}}ln \bigg(1+
{2\lambda \over nm C_\eta \mid \Lambda \mid R^2}(
R-\sum_{\sigma=1}^2 \sum_{i=1}^n \sum_{ j=1}^m
\parallel \phi_{ij}^\si \parallel_\infty)\bigg)\pb
\end{equation}
$\, \bull$

\sm
\begin{lemma} Let $T$ given by (\ref{cc1}) with $\lambda = R
C_\eta \mid \Lambda \mid$ $(R\geq R_1)$. Then then there exists
$\phi_0, R_2$ such that if $\parallel\phi  \parallel_1 \leq
\phi_0$, $ R \geq R_2$ one has
\begin{equation}\label{ee}
(\lambda T -1)^2 \geq {4C\over \lambda} \parallel \phi
\parallel_1 \vb
\end{equation}
where $C=Max(C_1,C_2)$ is given by (\ref{C1}) and (\ref{C2}).
\end{lemma}

\sm \noindent {\bf Proof  of Lemma 4.4:} Using the inequality $ln(1+x)\leq x, x
\geq 0$, then  one has:
\begin{equation}
\lambda T \nonumber \leq  { \lambda\over nm C_\eta  \mid \Lambda
\mid  R} ={1\over nm} \va
\end{equation}
which yields
\begin{equation} \label{T}
(\lambda T -1)^2 \geq \bigg({nm-1\over nm}\bigg)^2 \pa
\end{equation}
Then, if $\phi$ and  $ R $ are  such that
\begin{equation} \parallel \phi
\parallel_1 \leq {(nm-1)^2\over  8(nm)^3   }{C_\eta \mid \Lambda \mid\over (C_\eta  L_\mathcal{A}+ L_\eta)}= \phi_0,\end{equation}
and
\begin{equation}  \label{dd} R \geq {8 \over \mid \Lambda \mid}
\parallel \phi\parallel_1 {(nm)^2 \over  (nm-1)^2}=R_2.
\end{equation}
Then  (\ref{ee}) is proved. This completes the proof.
$\, \bull$

\sm Now we have all tools for the proof of the local existence theorem.

\sm \noindent{\bf Proof of Theorem 3.1:}  Let us consider  the following
subset in $X_T$ defined as follows:
\begin{eqnarray*} 
& & B_T  =\{ \psi=(\psi_{ij}^\si) \in X_T: \psi_{ij}^\si\geq 0, \parallel
\psi(t)\parallel_1 \leq a_0 \parallel \phi \parallel_1 , \,
\nonumber
\\ & &
\sum_{\sigma=1}^2
\sum_{i=1}^{n}\sum_{j=1}^{m}\widehat{\psi_{ij}^\si}(t,\, x-v_j
\cos(\th_i) t, y-v_j \sin(\th_i) t) \nonumber
\\ & &  \quad \leq R  \exp(\lambda t),
 t\in [0,T], x\in
\Omega\}.
\end{eqnarray*}

Let $\psi^1, \psi^2  \in B_T$, and let $R \geq R_0 =Max(R_1, R_2)$
then from Lemma 4.3, there exists $T$ such that
$\widehat{(A(\psi^1))_{ij }^\si}\geq 0$ and satisfying (\ref{es2})
for $\lambda =R C_\eta \mid \Lambda \mid$, $t\in [0,T]$. Moreover
from Lemmas 4.1 and 4.3,  one has:
\begin{equation}\label{bel}
\parallel A(\psi^1) \parallel_{X_T}\leq \parallel
\phi
\parallel_1 +{1  \over \lambda} C a_0^2 \parallel  \phi\parallel_1^2 +\lambda a _0 T
 \parallel \phi \parallel_1,
 \end{equation}
 and
 \begin{equation}\label{NI}
 \parallel A(\psi^1)- A(\psi^2)\parallel_{X_T}\leq
\bigg({2  \over \lambda } C a_0 \parallel \phi
\parallel_1 +\lambda T \bigg) \parallel \psi^1-\psi^2
\parallel_{X_T},
\end{equation}
where $C$  is the constants defined in Lemma 4.4.  Let
\begin{equation}\label{ww}
\Delta_0= (\lambda T -1)^2 - {4C \over \lambda}
\parallel \phi \parallel_1.
\end{equation}

 \noindent Then by the previous Lemma one has
$\Delta_0\geq 0$, for $\parallel \phi \parallel_1 \leq \phi_0$.
Let now $a_0$ be the positive quantity  given by:
\begin{equation} \label{kh} a_0= \lambda \,{(1- \lambda T) - \sqrt{\Delta_0}\over 2 C \parallel
\phi
\parallel_1}\va
\end{equation}
which  is a solution of
$$ \parallel \phi \parallel_1
+{1 \over \lambda } C a_0^2 \parallel \phi \parallel_1^2 +\lambda
a _0 T
\parallel \phi \parallel_1 = a_0  \parallel \phi
\parallel_1 \pb
$$
That shows from (\ref{bel}) that $A(\psi^1)\in B_T$. Moreover
$$ \bigg({2 \over \lambda }\, C \, a_0 \parallel \phi
\parallel_1  +\lambda \, T \bigg)= 1-\sqrt{\Delta_0}<1
$$
which shows  from (\ref{NI}) that the mapping $A$ is a contraction
in a ball $B_T$, and  application of   the fixed point theorem
gives the proof of Theorem 3.1, which refers to local existence.
Moreover, if  $\phi$ satisfies (\ref{bound2}), and  if $\parallel
\phi \parallel_1$ is very small, then from (\ref{cc}) and
(\ref{dd}), one can choose $R$ such that (\ref{bound3}) can be
obtained. This completes the proof. $\, \bull$


\vskip0.5cm

\section{Proof of  Existence for Large times}

The aim of this section consists in proving Theorem 3.2 concerning
the global existence of solutions to the initial value problem for
Eqs.(\ref{SM1})-(\ref{ini}). It will be proved that the solution
can be extended in each interval $[0, pT]$, for $ p \in N$. As in
the preceding section some preliminary results are needed:

\me \begin{lemma} Let $\phi$ satisfy the conditions of Theorem
3.1. Consequently, there exists  $\phi^1$ such that if $
\parallel \phi \parallel_1 \leq \phi^1$, there exists
$a_1$ such  the solution to Eqs.(\ref{SM1})-(\ref{ini}) can be
extended in the interval $[T, 2T ]$ and satisfies the estimate:
\begin{equation} \label{tt}\parallel \psi(t+T)\parallel_1 \leq a_1
\parallel \phi
\parallel_1, \quad t\in[0,T],
\end{equation}
\begin{equation} \label{bor}
 \rho(t+T,\bx) \leq R,  \quad \forall t\in [0,T], \quad \bx \in\Omega .
 \end{equation}

\end{lemma}

\sm \noindent{\bf Proof of Lemma 5.1:} Let
\begin{equation}\label{abc}
\Delta_1 = (\lambda T -1)^2 - {4\over \lambda}\, C \, a_0 \parallel \phi
\parallel_1,
 \end{equation}
where $ a_0, T$, and $\lambda$ are fixed and depending on $\phi_0$
and $R_0$. Then for
\begin{equation}
\parallel \phi
\parallel_1\leq  \phi^1= Min\bigg( \phi_0, {\lambda (\lambda T-1)^2  \over 4
Ca_0 }\bigg)
\end{equation}
the following $\Delta_1 \geq 0$ holds true.

We solve the problem (\ref{SM1})-(\ref{ini}) in $[T,2T]$ with initial condition given
by $\psi(T ; \bx)$. Then, for any $t \in [0,T]$, one has:
\begin{eqnarray}
 \widehat{\psi_{ij}^\si}(t+T,\bx)&=& \widehat{\psi_{ij}^\si}(T,\bx) + \int_0^{t}
\bigg(\widehat{\Gamma_{ij}^\si[\psi,\psi]}(s+T,\bx)\exp(-\lambda
(s+T)) \nonumber
\\ & + &\displaystyle \widehat{\psi_{ij}^\si}(s+T,\,\bx)\bigg\{\lambda - \widehat{ \Lambda
[\psi]}(s+T,\,\bx)\exp(-\lambda (s+T))\bigg\}\bigg)ds \pb
\end{eqnarray}

Consider now the ball
\begin{eqnarray}& & B^1_T =\bigg\{ g(t)=\psi(t+T) \in X_T: \psi\geq 0, \parallel
\psi(t+T)
\parallel_1 \leq a_1 \parallel \phi \parallel_1,
\nonumber \\ &&
\sum_{\sigma=1}^2\sum_{i=1}^{n}\sum_{j=1}^{m}\widehat{\psi_{ij}^\si}(t+T,\,
x-v_j \cos(\th_i) (t+T), y-v_j \sin(\th_i) (t+T))\nonumber \\ &
&\leq R \exp(\lambda (t+T)), \quad  t\in [0,T], \bx\in \Omega\
\bigg\}\nonumber\end{eqnarray}

Using the same technique as in the proof of Theorem 3.1 yields:
\begin{equation}\parallel A(\psi^1)(t+T) \parallel_{X_T}\leq a_0 \parallel
\phi
\parallel_1 +{1  \over \lambda} C a_1^2 \parallel  \phi\parallel_1^2 +\lambda a _1 T
 \parallel \phi \parallel_1,
\end{equation}
\begin{equation}\label{nie}\parallel A(\psi^1)- A(\psi^2)\parallel_{X_T}\leq
\bigg({2  \over \lambda } C a_1 \parallel \phi
\parallel_1 +\lambda T \bigg) \parallel \psi^1-\psi^2
\parallel_{X_T},
\end{equation}
when  $\psi^1, \psi^2 \in B^1_T$. By the same arguments as in the
proof of Theorem 3.1, and by choosing $a_1$ given by
\begin{equation} a_1= \lambda {(1- \lambda T) -
\sqrt{\Delta_1}\over 2 C \parallel \phi
\parallel_1} \geq 0 ,\end{equation}
the fixed point Theorem yields existence of solution in $[T, 2T]$.
This solution is continued in $[T, 2T]$ and in particular,
satisfies (\ref{tt}) and (\ref{bor}). This completes the proof of
Lemma 5.1.   $\, \bull$

\sm \noindent{\bf Proof of Theorem 3.2:} The iteration process can now be applied to prove global
existence in $[0, \infty[$. Suppose that for small initial data,
the solution exists and is continued $[0, (p-1)T]$ satisfying:

$$\parallel \psi(t+(r-1)T)\parallel_1 \leq a_{r-1}\parallel \phi
\parallel_1 , \quad r=1,....p-1, \quad t\in [0,T]\vb$$
and
$$\parallel \psi(r T)\parallel_1 \leq a_{r-1}\parallel \phi
\parallel_1 , \quad r=1,....p-1 \vb$$
where      $a_r$ are given by the following:
\begin{equation}  a_r= \lambda {(1- \lambda T) - \sqrt{\Delta_r}\over
2 C \parallel \phi
\parallel_1}, \qquad r=1,....p-1,  \end{equation}
\begin{equation} \Delta_r = (\lambda T -1)^2 -4C a_{r-1} {\parallel
\phi
\parallel_1 \over \lambda},\qquad  r=1,...p-1.\end{equation}
and the  reels $\Delta_0$ and $a_0$  are given respectively by
(\ref{ww}) and (\ref{kh}).

It can now be proved that the solution  can be extend in $[(p-1)T,pT]$,
satisfying for any $t\in [0,T]$ the following:
\begin{equation} \label{et1}\parallel \psi(t+(p-1)T)\parallel_1 \leq
a_{p-1}\parallel \phi
\parallel_1 \vb
\end{equation}
\begin{equation} \label{et2}\parallel \psi(p T)\parallel_1 \leq a_{p-1}\parallel \phi \parallel_1
\pb
\end{equation}

 Let $\psi=(\psi_{ij}^\si)$ be the solution of the following problem:
\begin{eqnarray}
 && \widehat{\psi_{ij}^\si}(t+(p-1)T,\bx) = \widehat{\psi_{ij}^\si}((p-1)T,\bx)\nonumber
\\ && \quad + \int_{(p-1)T}^{t+(p-1)T} \bigg(\widehat{\Gamma_{ij}^\si[\psi,\psi]}(s,\bx)\exp(-\lambda s)
 \displaystyle \widehat{\psi_{ij}^\si}(s,\,\bx)\nonumber
\\ && \quad\times \bigg\{\lambda - \widehat{
L[\psi]}(s,\,\bx) \exp(-\lambda s)\bigg\}\bigg)ds.
\end{eqnarray}

\noindent The same arguments as in the proof of Theorem 3.1 yields
\begin{equation}
\parallel A(\psi^1)(t+(p-1)T) \parallel_{X_T}\leq a_{p-2} \parallel  \phi
\parallel_1 +{ C a_{p-1}^2 \over \lambda} \parallel \phi \parallel_1^2
+\lambda a_{p-1} T \parallel \phi \parallel_1,
\end{equation}
and
\begin{eqnarray}& & \parallel A(\psi^1)(t+(p-1)T)-
A(\psi^2)(t+(p-1)T)\parallel_{X_T} \nonumber \\
&& \qquad \leq \bigg({2  C a_{p-1} \over \lambda} \nonumber \\
&& \parallel \phi \parallel_1  +\lambda T \bigg)\parallel \psi^1(t+(p-1)T)-\psi^2(t+(p-1)T)
\parallel_{X_T}\pb
\end{eqnarray}
The proof is complete if we choose  $a_{p-1}$ such that
\begin{equation}  a_{p-1}= \lambda {{(1- \lambda T) - \sqrt{\Delta_{p-1}}}\over
{2 C\parallel \phi \parallel_1}}\va
\end{equation}
where
\begin{equation}
\Delta_{p-1} = (\lambda T -1)^2 - {{4C a_{p-2}}  \over \lambda}
\parallel \phi
\parallel_1 \vb
\end{equation}
which give the  solution in $[(p-1)T,pT]$, satisfying
(\ref{et1})-(\ref{et2}). Then the proof is completed. $\, \bull$


\vskip.5truecm

\section{Simulations}

This section shows how numerical solutions can be obtained for the initial value problem in unbounded domains, namely within the framework of the qualitative analysis of Sections 4 and 5. An important technical difficulty is the computation of the left hand side of the equation modeling the transport term. The approach developed in this paper is based on the so-called splitting method, where the equation is splitted into the transport part and the interaction  term. High resolution finite differences methods for hyperbolic equations \cite{[GOSSE12]} are used.

In this paper, we propose to solve the system of partial differential equations by using a splitting method. As stated in \cite{[HOLD10],[Yazici2010]}, the idea behind this type of approach is that the overall evolution operator is formally written as a sum of evolution operators for each term in the model. In other words, one splits the model into a set of sub-equations, where each sub-equation is of a type for which simpler and more practical algorithms are available. The overall numerical method is then formed by picking an appropriate numerical scheme for each sub-equation and piecing the schemes together by operator splitting.

Let us suppose that the movement occurs with a unique velocity modulus $v$. Thus, this assumption of homogeneous velocity modulus let us introduce the notation $f_i^\si$, instead of $f_{ij}^\si$. After discretizing the velocity directions, Eq. (\ref{eq_gral}) reads:
\begin{equation} \label{splitting1}
\partial_t f_i^\sigma(t,\bx) + v\, \cos(\theta_i)\, \partial_x f_i^\sigma(t,\bx) + v\, \sin(\theta_i)\, \partial_y f_i^\sigma(t,\bx) = \mathcal{J}[\f](t,\bx).
\end{equation}

Then, we can  split this equation into the following set of sub-equations:
 \begin{equation} \label{splitting_dx}
 \partial_t f_i^\sigma(t,\bx) + v\, \cos(\theta_i)\, \partial_x f_i^\sigma(t,\bx) = 0,
 \end{equation}
\begin{equation} \label{splitting_dy}
 \partial_t f_i^\sigma(t,\bx) + v\, \sin(\theta_i)\, \partial_y f_i^\sigma(t,\bx) = 0,
 \end{equation}
and
\begin{equation} \label{splitting_dt}
 \partial_t f_i^\sigma(t,\bx) = \mathcal{J}[\f](t,\bx).
 \end{equation}

Notice that Eqs. (\ref{splitting_dx})-(\ref{splitting_dy}) are one-dimensional homogeneous transport equations, while Eq. (\ref{splitting_dt}) is just an ordinary temporal differential equation.  We will denote by $\mathcal{S}_x$, $\mathcal{S}_y$ and $\mathcal{S}_t$ the solution operators of Eqs. (\ref{splitting_dx})-(\ref{splitting_dt}), respectively. For instance, $\mathcal{S}_x u_0$ is the solution of Eq. (\ref{splitting_dx}) for an initial condition $u_0(\bx)$.

It is worth mentioning that one of the main advantages of this method is that the three new sub-equations can be properly solved by using a finite difference scheme. With that purpose, we discretize the time interval $[0,T]$ by taking $N_t$ equally spaced points $t_k = k \Delta t$, $k = 0, \ldots, N_t-1$. Now, given an initial condition $f_{i}^0 = f_i(0,\bx)$ for Eq. (\ref{splitting1}), we propose the following splitting process:

\sm \no
\textbf{Algorithm}

\textbf{Input:} $f_i^0$.

For $k=0,\ldots,N_t-2$ perform the following steps:
\begin{itemize}

\item[-] Solve Equation (\ref{splitting_dx}) with initial condition $f_{i}^{k}$, obtaining a solution $\mathcal{S}_x f_{i}^{k}$.

\vspace*{0.3cm}
\item[-] Solve Equation (\ref{splitting_dy}) with initial condition $\mathcal{S}_x f_{i}^{k}$, obtaining a solution $\mathcal{S}_y \mathcal{S}_x f_{i}^{k}$.

\vspace*{0.3cm}
\item[-] Solve Equation (\ref{splitting_dt}) with initial condition $\mathcal{S}_y \mathcal{S}_x f_{i}^{k}$, obtaining a solution $\mathcal{S}_t \mathcal{S}_y \mathcal{S}_x f_{i}^{k}$.

\vspace*{0.3cm}
\item[-] Set $f_{i}^{k+1} = \mathcal{S}_t \mathcal{S}_y \mathcal{S}_x f_{i}^{k}$.
\end{itemize}

\textbf{Output:} $f_i^k$, $k=0,\ldots,N_t-1$.
\sm

The output of the algorithm is the numerical solution of the original problem (\ref{splitting1}), given by $f_i^\sigma(k\Delta t, \bx) = f_i^k, k = 0, \ldots, N_t-1$, while the interested reader is referred to \cite{[HOLD10]} for a detailed analysis of the convergence of the method under consideration.

In the next paragraphs we propose two case-studies in order to carry out some simulations, to analyze the accuracy of the model and of the proposed numerical scheme and to investigate the emerging behaviors.

\subsection*{Case-study 1}
\begin{figure}
\begin{center}
\includegraphics[scale=0.4]{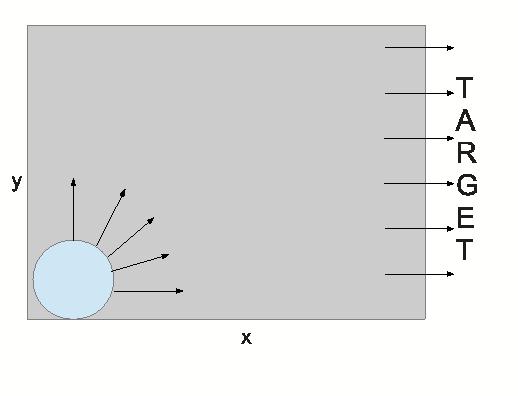}
\caption{Initial condition for case-study 1. A crowd localized in the bottom-left corner is constituted by individuals moving in five different directions, indicated by the arrows. The parallel arrows on the right show the exit of the room.} \label{case_study1_IC}
\end{center}
\end{figure}

Let us first consider a group of individuals initially located near the corner of a rectangular room represented by the domain $\Omega = [0,1]\times [0,1] \subseteq {\RR}^2$, as shown in Figure \ref{case_study1_IC}.

Let us suppose that all of them are moving, at time $t=0$, with constant velocity $v$ but with five different directions given by $\theta_i =  i \pi /8$, $i = 0, \ldots, 4$. Moreover, the individuals are equally distributed among these directions. In that moment some panic situation occurs at that corner and all of them try to move towards the exit, which is indicated by the parallel arrows in Figure \ref{case_study1_IC}.

\begin{figure}
\begin{tabular}{cc}
\includegraphics[width=0.46\textwidth]{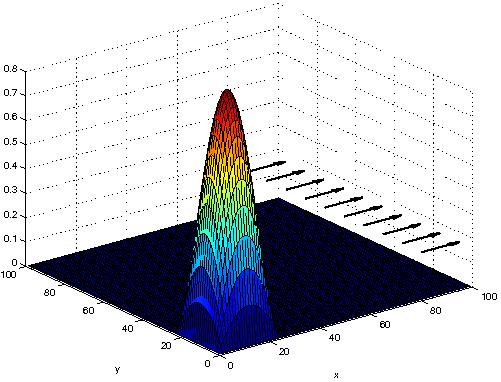} &
\includegraphics[width=0.46\textwidth]{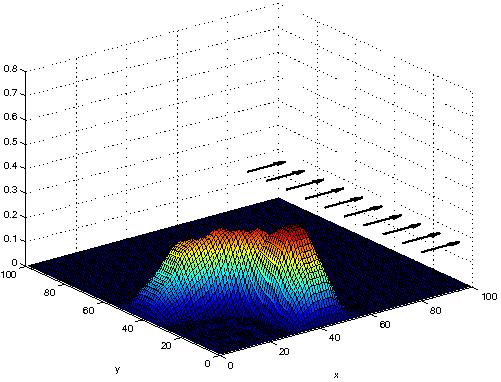}
\\
(a) & (b)
\\
\includegraphics[width=0.46\textwidth]{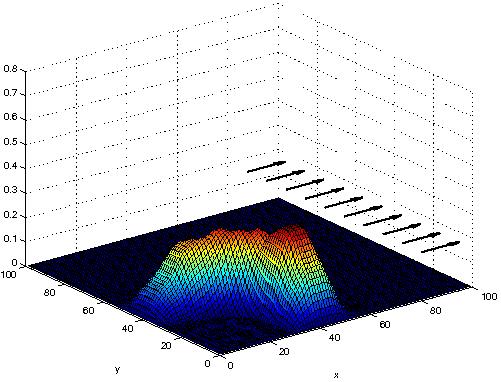} &
\includegraphics[width=0.46\textwidth]{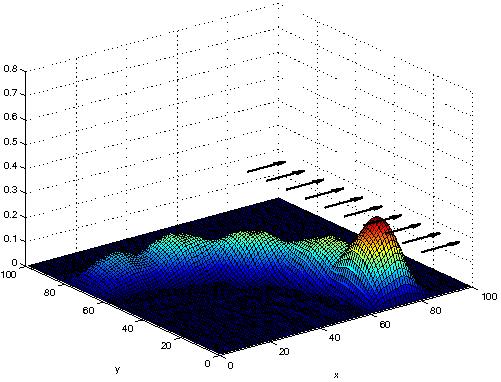}
\\
(c) & (d)
\end{tabular}
\caption{Case-study 1. Evolution of the density of the crowd for different times. (a) corresponds to the initial density, while (b)-(d) show its evolution. The parallel arrows indicate the target.}
\label{case_study1_evolution}
\end{figure}

Notice that those individuals who are moving in the direction $\theta_0 = 0$ are the most favored, as they are already
moving towards the exit. However, we assume that the other individuals start to change their directions according to a
table of games considering only the sensitivity to reach the target. Assuming that the activity is homogeneously
distributed, i.e. $u_0$ is constant, that $\Lambda(\bx)=\bx$, and taking $v = 0.03$ and $\alpha_1 = \alpha_2 = \alpha =
0.06$ we perform some simulations, obtaining the results shown in Figure \ref{case_study1_evolution}.

We can observe that as time evolves the pedestrians are closer to the exit, because many of them were able to change their directions. It is worth mentioning that the fact of considering only the tendency to reach the target let the individuals correct their movements without getting confused by the tendency to follow the stream.

\begin{figure}
\begin{tabular}{cc}
\includegraphics[width=0.46\textwidth]{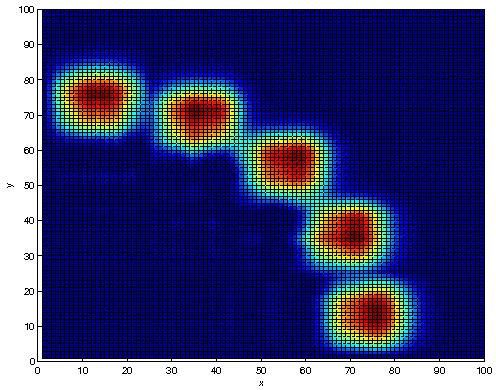} &
\includegraphics[width=0.46\textwidth]{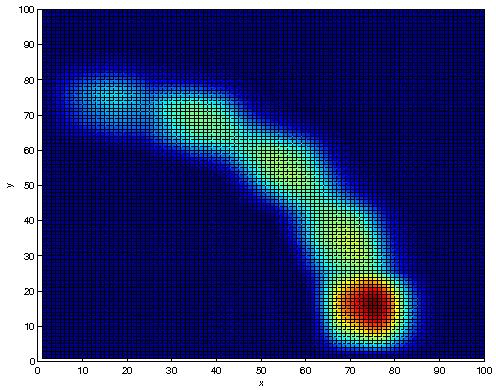}
\\
(a) & (b)
\end{tabular}
\caption{Case-study 1. Comparison between the evolution for (a) $\alpha = 0$ and (b) $\alpha = 0.06$. }
\label{case_study1_comparison}
\end{figure}

In Figure \ref{case_study1_comparison} we show a top view comparing the cases $\alpha = 0$ and $\alpha = 0.06$ for a fixed time. It becomes clear that originally the five groups were following different targets and that the model is successful predicting the reaction after a panic situation.

\subsection*{Case-study 2}

\begin{figure}
\begin{center}
\includegraphics[scale=0.4]{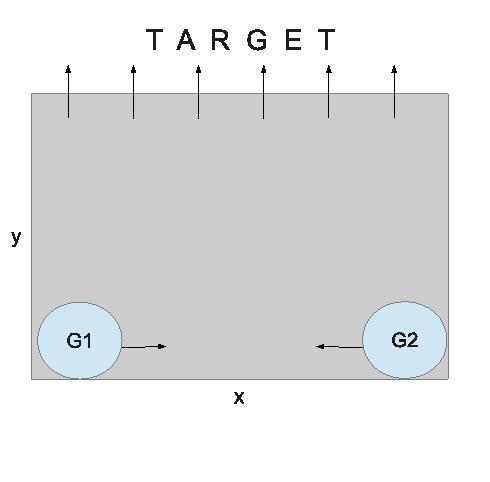}
\caption{Initial condition for case-study 2. Two crowds originally moving in opposite directions, indicated by the arrows. The parallel arrows in the top show the exit of the room.}
\label{case_study2_IC}
\end{center}
\end{figure}

Let us consider two different groups, labeled by $G1$ and $G2$, which are initially moving towards two different targets inside the rectangular room, as shown in Figure \ref{case_study2_IC}. Both groups have the same velocity modulus $v$ and we consider five directions of movement $\theta_i = i \pi /4$, $i = 0, \ldots, 4$. In this way, $G1$ and $G2$ are initially moving in the directions $\theta_0 = 0$ and $\theta_4=\pi$, respectively.

Once more, we assume that they realize about the location of the exit, which corresponds to the direction $\theta_2 = \pi /2$, and they consequently try to adapt their movement. Notice that in this case none of the groups originally have the proper direction of movement.

\begin{figure}
\begin{tabular}{cc}
\includegraphics[width=0.46\textwidth]{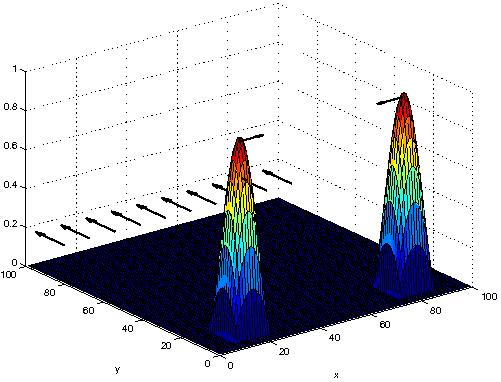} &
\includegraphics[width=0.46\textwidth]{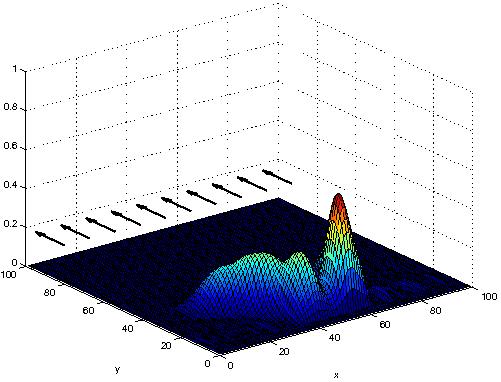}
\\
(a) & (b)
\\
\includegraphics[width=0.46\textwidth]{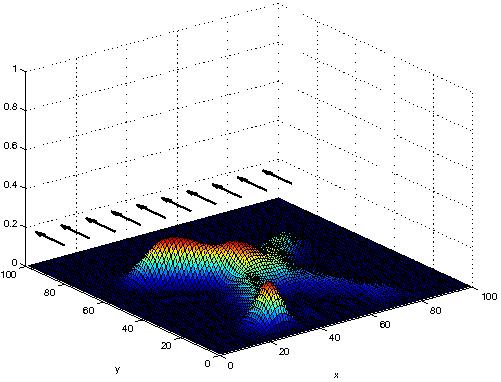} &
\includegraphics[width=0.46\textwidth]{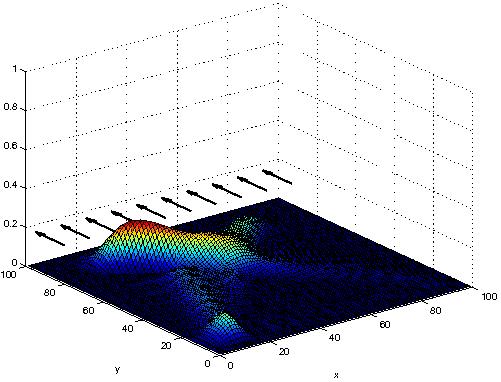}
\\
(c) & (d)
\end{tabular}
\caption{Case-study 2. Evolution of the density of the crowds for different times. (a) corresponds to the initial density and the arrows on each crowd indicate the initial direction, while (b)-(d) show the evolution. The parallel arrows indicate the target.}
\label{case_study2_evolution}
\end{figure}

The table of games takes into account that the individuals can change their direction with a certain probability in order to reach the exit. However, we consider that individuals from group $G1$ have better abilities, changing their direction with a probability $\alpha_1 = 0.2$, while individuals from group $G2$ do it with a probability $\alpha_2 = 0.1$. The simulations are performed with these values and taking $\Lambda(\bx)=\bx$ and $v = 0.03$. The results are shown in Figure \ref{case_study2_evolution}.

\begin{figure}
\begin{tabular}{cc}
\includegraphics[width=0.46\textwidth]{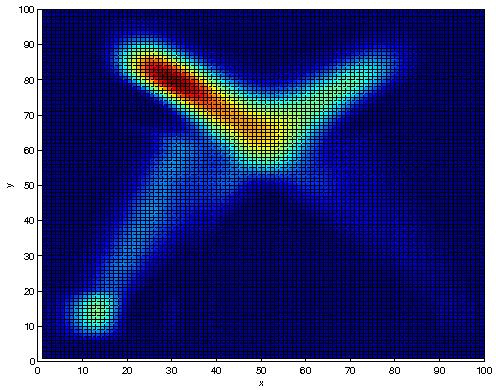} &
\includegraphics[width=0.46\textwidth]{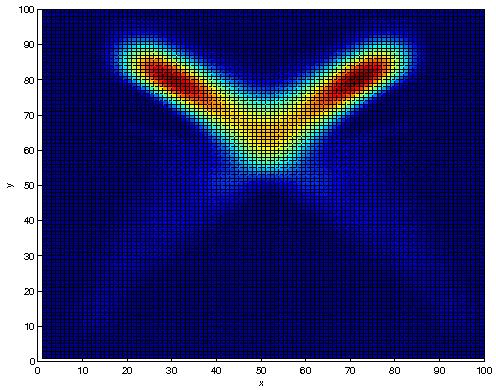}
\\
(a) & (b)
\end{tabular}
\caption{Case-study 2. Comparison between the evolution for (a) $\alpha_1 = 0.2$, $\alpha_2 = 0.1$ and (b) $\alpha_1 = \alpha_2 = 0.2$. }
\label{case_study2_comparison}
\end{figure}

In Figure \ref{case_study2_comparison} we show a top view comparing the afore-said situation with the case in which $\alpha_1 = \alpha_2 = 0.2$. Notice about the asymmetric shape of the crowd in Figure \ref{case_study2_comparison}(a), clearly given by the different abilities (e.g. visual, fitness) that each group exhibits.

\vskip.5truecm
\section{Looking Ahead to Perspectives}

Various approaches to modeling crowd dynamics  corresponding to different representation scales and related mathematical tools have been reviewed and critically analyzed in the survey paper \cite{[BPT12]}. The mesoscopic approach in this present paper has been selected, out of this overall analysis, as the most appropriate, according to the authors' bias, to capture the complexity features of the system under consideration. Moreover, the use of discrete velocity variables has been adopted to model the micro-scale state of pedestrians in order to capture the granular features of their dynamics \cite{[CPT11]}.

The use of discrete velocity variables was introduced, as already mentioned, to model vehicular traffic\cite{[DT07],[CDF07]}, while the approach was further developed toward crowd modeling in \cite{[BB11]}. The discretization is here applied to the
velocity variable. In principles, it can be referred also to the space variable as already known for models of vehicle dynamics \cite{[FT12]}. However, the technical difficulty of dealing with more than one space variables creates nontrivial technical problems, which do not seem yet solved at present.

The modeling approach proposed in this paper refers to normal flow conditions, namely when  pedestrians are sufficiently relaxed to behave normally. On the other hand, the behavioral influence of panic conditions can play an important role in the modification of the usual dynamics
\cite{[HFV00],[HJA07]}. This objective can be pursued by an appropriate tuning of the model proposed in this paper also by taking advantage of recent studies on the behavioral dynamics of pedestrians \cite{[MHG09],[RWTH11]}, and evolutive game theory \cite{[CFB11]}. Accordingly, the guidelines
toward this important objective are given:

\begin{itemize}

\item Pedestrians undergo a trend to higher values of the activity variable. Moreover, the dynamics over the said variable has to be necessarily inserted in the model;

\item The effective visibility zone becomes larger and signals from large distance become important, while in the case of normal conditions short distance signals appear to be more important;

\item Pedestrians appear to be sensitive to crowd concentration,  while  in  normal conditions they attempt to avoid it. Therefore different weights $\varepsilon$ need to be used;

\item The role of boundary conditions needs to be properly modeled arguably by means of nonlocal conditions

\end{itemize}

Moreover an important topic to be studied is the role of control actions by external systems. Namely the system needs to be modeled as an open one.


\end{document}